\documentclass[10pt,letterpaper]{article}
\usepackage{amsmath,amssymb}
\usepackage{changepage}
\usepackage[utf8x]{inputenc}
\usepackage{textcomp,marvosym}
\usepackage{cite}
\usepackage{nameref,hyperref,graphicx}
\usepackage[right]{lineno}
\usepackage{microtype}
\DisableLigatures[f]{encoding = *, family = * }
\usepackage[table]{xcolor}
\usepackage{array}

\usepackage{caption}
\graphicspath{{Figures/}} 
\usepackage{authblk}
\usepackage[margin=1in]{geometry}

\title{Coupled Contagion: A Two-Fears Epidemic Model}
\author[1]{Joshua M. Epstein}
\author[1]{Erez Hatna}
\author[2]{Jennifer Crodelle}
\affil[1]{Department of Epidemiology, School of Global Public Health, New York University, New York, NY, USA}
\affil[2]{Department of Mathematics, Middlebury College, Middlebury, VT, USA}
\date{}                     %% if you don't need date to appear
\begin{document}
  \maketitle

%%%%%%%%%%%%%%%%%%%%%%%%%%%%%%%%%%%%%%%%%%%%%%%%%%%%%%%%%%%%%%%%%%%
%%%%%%%%%%%%%%%%%%%%%%%  ABSTRACT %%%%%%%%%%%%%%%%%%%%%%%%%%%%%%%%
%%%%%%%%%%%%%%%%%%%%%%%%%%%%%%%%%%%%%%%%%%%%%%%%%%%%%%%%%%%%%%%%
\begin{abstract}
We present a differential equations model in which contagious disease transmission is affected by contagious fear of the disease and contagious fear of the control, in this case vaccine. The three contagions are coupled. The two fears evolve and interact in ways that shape distancing behavior, vaccine uptake, and their relaxation. These behavioral dynamics in turn can amplify or suppress disease transmission, which feeds back to affect behavior.  The model reveals several coupled contagion mechanisms for multiple epidemic waves. Methodologically, the paper advances infectious disease modeling by including human behavioral adaptation, drawing on the neuroscience of fear learning, extinction, and transmission.
\end{abstract}

%%%%%%%%%%%%%%%%%%%%%%%%%%%%%%%%%%%%%%%%%%%%%%%%%%%%%%%%%%%%%%%%%%%
%%%%%%%%%%%%%%%%%%%%%% %% INTRODUCTION %%%%%%%%%%%%%%%%%%%%%%%%%%%%%%%%
%%%%%%%%%%%%%%%%%%%%%%%%%%%%%%%%%%%%%%%%%%%%%%%%%%%%%%%%%%%%%%%%
\section{Introduction}\label{intro}

In classical mathematical epidemiology -- the venerable tradition of the 1927 Kermack-McKendrick model -- individuals do not adapt their contact behavior during epidemics \cite{Kermack1927, brauer2005}. Specifically, they do not endogenously engage in social distancing based on fear. Yet, such behavior is well-documented in true epidemics. In 2008, Epstein et al. published \emph{Coupled Contagion Dynamics of Fear and Disease} \cite{epstein2008a}, a model that introduced the idea of two interacting contagions: one of the disease proper and one of fear of the disease. The model’s core narrative is that epidemic growth induces fear. Contagious fear among healthy susceptible people, in turn, induces self-isolation. By depriving the epidemic of fuel, in the form of susceptibles, this self-isolation suppresses the disease. When disease prevalence becomes low, however, so does the fear. Thus, susceptible people (no longer fearful) come out of hiding. But, because there are still infectious individuals in circulation, this pours gasoline (susceptible individuals) on the remaining embers (the infectives), igniting a second wave. Precisely this occurred historically, in the 1918 Pandemic Flu (see \cite{bootsma2007}), and history repeated itself in the multi-wave COVID-19 pandemic  \cite{CDC2020}. Recent work on the neuroscience of fear lends scientific support to the postulate of fear contagion, and a recent agent-based model explicitly includes fear modules grounded in that neuroscience; see \cite{epstein2013, epstein2016}. The literature on behavioral adaptation in epidemics has grown in several directions; see \cite{fenichel2011a, funk2010, perra2011, weston2018}.

In the present work, we modify and extend the original coupled contagion model \cite{epstein2008a} in light of recent advances, subsuming it in a more general framework that—while including contagious fear of disease—adds contagious fear of vaccine. The World Health Organization recently included vaccine refusal in the top ten threats to global health \cite{medicine2019a}. It is responsible for the resurgence of several deadly vaccine-preventable diseases, including measles and pertussis in the US and even polio in several countries \cite{patel2019, dube2015}. During the swine flu pandemic of 2009, roughly 40 percent of Americans refused the vaccine \cite{blasi2012}. And, writing as COVID-19 vaccination is beginning, there is concern that refusal will undermine the attainment and maintenance of herd immunity to the SARS-CoV-2 virus and its variants.
 
%%%%%%%%%%%%%%%%%%%  CORE IDEA 1.1 %%%%%%%%%%%%%%%%%%%
 \subsection{The Core Idea}
In our model, as discussed in \cite{epstein2020}, ``Everything turns on the relationship between the two fears, one of disease, the other of vaccine." If fear of the disease exceeds fear of the vaccine in the population, the rate of vaccine acceptance rises, and the disease may be suppressed. However, if the prevalence of the disease is suppressed enough, fear of the disease may fall below fear of the vaccine (as might happen when a disease recedes from our collective memory). Now the vaccine is scarier than the disease, people eschew the vaccine, and a new disease cycle can explode.

This narrative also rings true historically. Smallpox, one of the great scourges of human history, kills roughly 30 percent of those infected  \cite{fenner1988}. Yet, even when inoculation (with cowpox) was discovered, cycles of vigilance and complacency kept smallpox alive. In her social history of smallpox, \emph{the Speckled Monster}, Jennifer Lee Carrell \cite{carrell2003} recounts, ``In London, inoculation's popularity waxed and waned through the 1730s, with the force of the disease: in bad years, people flocked to be inoculated; in lighter years, the practice shrank. Inoculation was a security—the only security—to cling to within the terror of an epidemic; in times of good health, however, it looked like a foolish flirtation with danger." Our two-fear model generates such cycles and related dynamics.  

%%%%%%%%%%%%%%%%%%% CORE IDEA 1.2 %%%%%%%%%%%%%%%%%%%
 \subsection{Irrational Epidemics}
 
 As discussed in \cite{epstein2008a}, our approach differs from the so-called ``rational epidemics” tradition (stemming from \cite{geoffard1996, kremer1996}) which does not model contagious fears. Rather, as in microeconomics and game theory, agents maximize utility conditional on the disease's prevalence. While illuminating in several important settings, prevalence-elastic optimal adaptation in the rational choice tradition is not well-suited to capture prevalence independent fear contagions such as Morgellan's disease, an internet-disseminated delusional parasitosis \cite{vila-rodriguez2008}, or the mass panics that occurred in Surat India (1994), or during Ebola \cite{towers2015}, or in recent episodes of vaccine refusal \cite{broniatowski2018}. Indeed, cognitive neuroscience demonstrates that the fear response and fear learning generally are not fundamentally choice-like, or even necessarily conscious, none of which means they can't be modeled or counteracted \cite{epstein2013, epstein2016, ledoux2000, ledoux2003a, ledoux2003b, ledoux2009a, ledoux2012a}.

 \subsection{Organization}
After the model's technical exposition (Section \ref{model}), we offer four base scenarios with discussions of their dynamics and importance (Section \ref{results}). Analytical results and extensive sensitivity analyses are provided and discussed in Section \ref{sensAnalysis}.

We begin with the pure compartmental Susceptible-Infected-Recovered (SIR) version of a contagious disease alone. Every subsequent scenario subsumes the preceding one, as follows:\\

\noindent Scenario 1: contagious disease\\
Scenario 2: contagious disease + fear of the disease\\ 
Scenario 3: contagious disease + fear of the disease + vaccination\\ 
Scenario 4: contagious disease + fear of the disease + vaccinations + fear of the vaccinations\\

These four scenarios are of central concern to public health.  All numerical assumptions (parameter settings and initial conditions) are given in the Supplementary Information (SI), ensuring replicability. Several mathematical conditions for growth are derived there as well. The SI also includes a pure fear ``Salem Witches" scenario, where fear propagates in the absence of any disease, further distinguishing the approach from prevalence elastic rational adaptation. On emotional contagion and its mechanisms, see \cite{epstein2013, hatfield1993}.
 
%These four scenarios are central to public health. In the , we also offer a pure fear scenario, where fear propagates in the absence of disease. 

%%%%%%%%%%%%%%%%%%%%%%%%%%%%%%%%%%%%%%%%%%%%%%%%%%%%%%%%%%%%%%%%%%%
%%%%%%%%%%%%%%%%%%% SECTION 2 - THE MODEL %%%%%%%%%%%%%%%%%%%%%%%%%%%%%%%%
%%%%%%%%%%%%%%%%%%%%%%%%%%%%%%%%%%%%%%%%%%%%%%%%%%%%%%%%%%%%%%%%

\section{The Model}\label{model}
Proceeding with the exposition, we first define all state variables and parameters of the model in Tables \ref{varDefs} and \ref{paramDefs} respectively. It may be of interest that we use an average infectious period of seven days ($\frac{1}{\gamma} = 7$) and a basic reproduction number ($R_0$) of two ($\frac{\beta}{\gamma} = 2$) for the scenarios discussed. % and \textcolor{blue}{an effective contact rate of X }. 

\begin{table}[h!]
\renewcommand{\arraystretch}{1.4}
\begin{tabular}{|c|l|}
\hline
\cellcolor[gray]{0.8}  \textbf{Variable} & \cellcolor[gray]{0.8}  \textbf{Description}\\
\hline 
$S(t)$ & The proportion of susceptible individuals with no fear \\
\hline
$S_{fd}(t)$ &  The proportion of susceptible individuals who fear the disease\\
\hline
$S_{fv}(t)$ &  The proportion of susceptible individuals who fear the vaccine\\
\hline
$I(t)$ &  The proportion of (pathogen) infectious individuals\\
\hline
$R_{nat}(t)$ &  The proportion of recovered individuals (persons who had the disease and gained immunity)\\
\hline 
$R_{vac}(t)$ &  The proportion vaccinated individuals\\
\hline
$A(t)$ &  The proportion of recently-vaccinated individuals who fear the vaccine because of an adverse reaction\\
\hline
$v(t)$ & The rate of vaccination (1/days)\\
\hline
\end{tabular}
\caption{State variable definitions.}
\label{varDefs}
\end{table}

\begin{table}[h!]
\renewcommand{\arraystretch}{1.4}
\begin{tabular}{|c|l|}
\hline
\cellcolor[gray]{0.8}  \textbf{Parameter} & \cellcolor[gray]{0.8}  \textbf{Description}\\
\hline 
$\beta$ & the effective contact rate for the pathogen (1/days)\\
\hline
$\beta_{fd}$ &  the effective contact rate of fear of the disease (1/days)\\
\hline
$\beta_{fv}$ &  the effective contact rate of vaccine fear (1/days)\\
\hline
$\alpha_{f}$ & the effective contact rate of fear loss (1/days)\\
\hline
$\gamma$ & the rate of disease recovery  (1/days) \\
\hline
$\gamma_f$ & the rate of spontaneous loss of fear (1/days) \\
\hline
$p$ & the relative risk of acquiring the disease for pathogen-fearful individuals\\
\hline
$\eta$ & the fear difference scaling factor \\
\hline
$\sigma$ &the fraction of the rate of vaccinated that experience adverse effects \\
\hline
$\epsilon$ & the maximum rate of vaccination (1/days)\\
\hline
\end{tabular}
\caption{Parameter identifications.}
\label{paramDefs}
\end{table}

\newpage

The mathematical model relating these variables and parameters consists of the eight coupled nonlinear ordinary differential equations shown below. For expository efficiency, we use a well-mixed model. Natural extensions would include social networks \cite{wang2015, newman2010, eubank2004, kiss2017} and agent-based formulations \cite{parker2011}.

\begin{align}
\frac{dS}{dt} &= -\beta IS   - \beta_{fd}(S_{fd} + I)S - \beta_{fv}(S_{fv}  +A)S + \gamma_f( S_{fd} + S_{fv}) + \alpha_{f}(R_{nat} S_{fd} + R_{vac}S_{fv})\label{diffeq1} \\
\frac{dS_{fd}}{dt} &= - p\beta  I S_{fd}  - \gamma_fS_{fd} - \alpha_{f}  R_{nat}S_{fd} + \beta_{fd}(S_{fd} + I)S - vS_{fd}\label{diffeq2} \\
\frac{dS_{fv}}{dt} &= -\beta I S_{fv}  - \gamma_f S_{fv} - \alpha_{f} R_{vac}S_{fv}   + \beta_{fv}(S_{fv}  +A)S\label{diffeq3}\\
\frac{dI}{dt} &= \beta IS + p\beta I S_{fd} + \beta I S_{fv} - \gamma I\label{diffeq4} \\
\frac{dR_{nat}}{dt} &= \gamma I\label{diffeq5}\\
\frac{dR_{vac}}{dt} &= (1-\sigma)vS_{fd} +  \gamma_f A\label{diffeq6} \\
\frac{dA}{dt} &= \sigma v S_{fd} - \gamma_f A \label{diffeq7}\\
\frac{dv}{dt} &= \eta(S_{fd} - S_{fv})v(\epsilon - v)\label{diffeq8},
\end{align}

\begin{figure}[h!]
\centering
\includegraphics[scale=0.32]{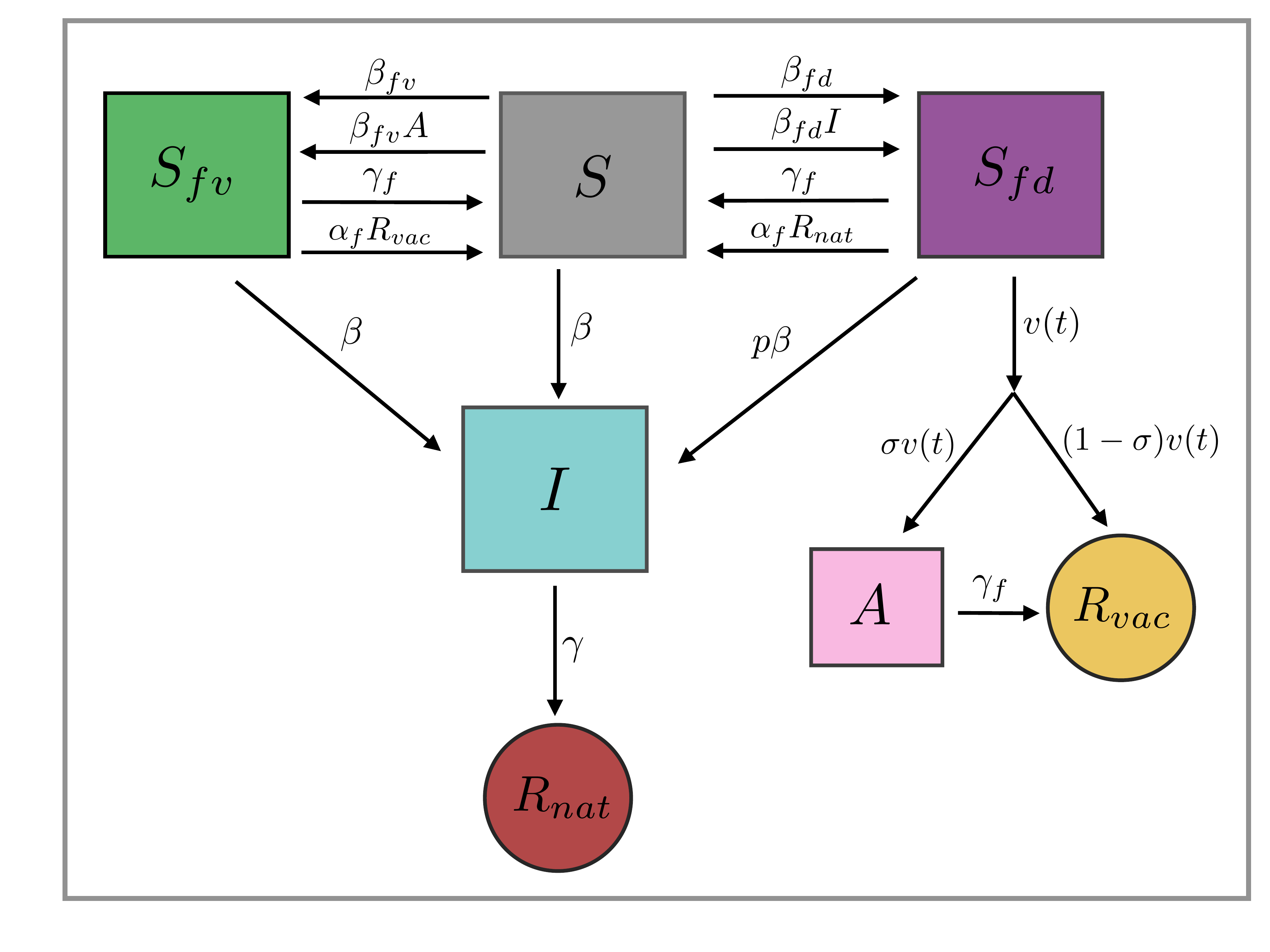}
\caption{Flow diagram of the equations (Eqs (\ref{diffeq1}) - (\ref{diffeq8})).}
\label{fig_schem}
\end{figure}

Capital letters indicate infection states, while subscripts indicate fear states. For example, the $S_{fd}$ compartment is the fraction of the population that is susceptible to the disease and fears the disease (subscript $fd$). The $S_{fv}$ compartment is the fraction of the population that is susceptible to the disease and fears the vaccine (subscript $fv$). While this dynamical system is rich, as discussed below, Eqs (\ref{diffeq1}), (\ref{diffeq4}), and (\ref{diffeq5}) reduce to the familiar SIR model when there are no fears and the terms subscripted by $fd$ or $fv$ are set to zero. Moreover, each of the fear contagions also propagates in classical fashion, as can be seen in Eqs (\ref{diffeq2}) and (\ref{diffeq3}). A flow diagram of the model is shown in Figure \ref{fig_schem}.   

%\clearpage %

%%%% FIGURE 1: SCHEMATIC %%%%%

\newpage
%%%%%%%%%%%%%%%%%%%   TRANSMISSION OF FEAR 2.1  %%%%%%%%%%%%%%%%%%%
\subsection{Transmission of disease and fear}
The equations include six population compartments, each representing the proportion of individuals in the given state at any time. The sum of these six compartments is always 1.  A susceptible individual may acquire the disease by being exposed to an infectious person. The effective contact parameter represents the rate of transmission. When effective contact occurs, a susceptible person becomes infectious for an average period of $\frac{1}{\gamma}$ days. Once the infectious period ends, the individual recovers and gains permanent immunity to the pathogen (compartment $R_{nat}$). We simplified the model by not including a pre-infectious (latent) period,  although this is an obvious extension.

We consider three types of susceptible individuals: persons without fear ($S$), persons who fear the disease ($S_{fd}$), and persons who fear the vaccine ($S_{fv}$). A susceptible person can retain only one fear at a given time, and all fearful persons share the same fear intensity.  A non-fearful (susceptible) individual may acquire fear of the disease by interacting with infectious or disease-fearful persons. These processes represent scenarios in which a susceptible person observes or communicates with an infectious (sick) individual or with a disease fearful person. Unlike the transmission of the pathogen, such interactions could occur at a distance (as on social media) and thus require a dedicated effective contact rate parameter ($\beta_{fd}$). Note that an infectious individual can infect a susceptible person with either the pathogen or fear of the disease, but again, not both.

%%%%%%%%%%%%%%%%%%%   FEAR OF DISEASE 2.2 %%%%%%%%%%%%%%%%%%%
\subsection{Fear of disease}
Fear of the disease affects the behavior of susceptible individuals. These persons may take protective actions, such as self-isolation, mask-wearing, social distancing, avoidance of travel and mass gatherings, and improved personal hygiene. In the interest of simplicity, such actions are modeled using the relative risk parameter $p$, which is used to scale down $\beta$. This is a fundamentally different representation than in \cite{epstein2008a}, where distanced individuals were a separate compartment. Here, they are not. In the present model, a value of $p = 0.25$ represents a 75\% decrease in the likelihood of a disease-fearful individual becoming infected with the disease compared to a susceptible individual with no fear. Disease-fearful individuals may also choose to gain permanent immunity through vaccination. We assume that a small proportion ($\sigma$) of vaccinated individuals experience adverse effects or associate an unrelated discomfort with the vaccine. These individuals ($A$) acquire a (transmissible) fear of the vaccine while gaining full immunity. The rest of the vaccinated individuals, a proportion of $1 -\sigma$, gain immunity without acquiring the fear ($R_{nat}$).

%%%%%%%%%%%%%%%%%%%   FEAR OF VACCINE 2.3 %%%%%%%%%%%%%%%%%%%
\subsection{Fear of the vaccine}
Susceptible individuals acquire fear of the vaccine by interacting with vaccinated persons who had an adverse experience ($A$) or with vaccine fearful susceptible persons ($S_{fv}$). The effective contact rate of such interactions is $\beta_{fv}$.

%%%%%%%%%%%%%%%%%%%   FEAR EXTINCTION 2.4 %%%%%%%%%%%%%%%%%%%
\subsection{Fear extinction}
We know from neuroscience that, post-traumatic stress notwithstanding, fear is not permanent but decays in the absence of an aversive stimulus. In this model, susceptible people may naturally overcome both fears (of disease and vaccine) and join the compartment of non-fearful susceptible individuals ($S$). Our model contains two paths for such fear decay, or ``extinction" as it is called in behavioral neuroscience \cite{norrholm2008}. Specfically, we think of exposures (direct or indirect) to disease-infected people as classical associative fear-conditioning trials. A classic example of a fear conditioning trial is as follows. If a person is simply shown a benign blue light, no manifestations of fear (e.g., freezing, pupil dilation, adrenaline spikes, increased heart rate, electrodermal activity) or neural correlates of fear, such as activation of (e.g., oxygenation and recruitment of blood to) the amygdala, as seen in fMRI \cite{fullana2016} are observed. By contrast, if the subject is unexpectedly given an aversive electric shock, the amygdala is immediately stimulated, triggering a suite of fear responses. Importantly, if the two stimuli are repeatedly paired -- blue light followed shortly by shock -- the subject will come to associate (not necessarily consciously) the light with the shock, to the point where the blue light alone elicits the same amygdala response as the shock. By a process of associative learning, the subject has been "conditioned" to fear the blue light. If these light-shock pairings are discontinued, the fear of the blue light will decay. Both the fear acquisition phase and fear extinction phases can be modeled mathematically \cite{gershman2010}.   

Consider a person whose fear of the disease has prompted self-isolation. This person's fear may decay in two ways. The first is by eliminating direct (aversive) exposures to disease-infected individuals; conditioning trials are thereby suspended, and ``fear extinction" commences. In the model, this natural decay is exponential, consistent with the simple seminal Rescorla-Wagner model \cite{rescorla1972}. We assume that, in the absence of a fear stimulus, a person will retain fear for an average duration of $\frac{1}{\gamma_f}$ days. On the widespread use of the Rescorla-Wagner model, see \cite{siegel1996}. For other learning models, see for example \cite{pearce1980} and \cite{sutton1998}.

The second path to overcoming fear is social and distinct from extinction through stimulus deprivation. Individuals may lose fear by communicating with persons who have recovered from the fearful event. These reassuring exposures (think of repeated blue light and candy pairings) can damp the conditioned fear. This would be called counter-conditioning, over-writing a negative response with a positive one. (On the relative effectiveness of extinction and counter-conditioning in diminishing children's fear, see \cite{newall2017}). By interacting with a recovered person ($R_{nat}$), a disease fearful person ($S_{fd}$) may lose their fear. Similarly, a vaccine-fearful person ($S_{fv}$) may lose their fear by interacting with a protected vaccinated person ($R_{vac}$). 

In our model, vaccinated persons who had gained fear due to a negative vaccine experience ($A$) abandon the fear only via the first path: natural exponential decay. We assume that their first-hand experience with the vaccine makes them resistant to social influence. Analogous to disease fear, they retain their fear of vaccine for an average duration of $\frac{1}{\gamma_f}$ days and then join the compartment of vaccinated individuals ($R_{vac}$).

Widespread distancing and vaccination also cut the disease's growth rate and can even make it negative--the herd immunity condition--which amplifies their suppressive effects.

%%%%%%%%%%%%%%%%%%%   VACCINE UPTAKE 2.5 %%%%%%%%%%%%%%%%%%%
\subsection{Vaccine uptake}
The daily rate at which fearful disease-susceptible persons vaccinate, $v(t)$, may change over time due to a mechanism of social influence; see Eq (\ref{diffeq8}). Specifically, we assume that the growth rate of $v(t)$ increases $\left(\frac{dv}{dt} > 0\right)$ when fear of the disease exceeds the fear of the vaccine. It decreases $\left(\frac{dv}{dt} < 0\right)$ when the reverse obtains -- when vaccine-fear is more prevalent than disease-fear. We represent this effect using the difference between the two prevalences ($S_{fd} - S_{fv}$):

\[ \frac{dv}{dt} = \eta(S_{fd} - S_{fv})v(\epsilon - v) \]

Clearly, $S_{fd} - S_{fv} = 0$, is a tipping point of the dynamics. Several mechanisms can affect the fear ordering. If the model begins with disease fear exceeding vaccine fear ($S_{fd} - S_{fv} > 0$), vaccination expands. However, this itself can endogenously suppress the disease to the point where fear of disease falls below fear of vaccine. At this point, the fear ordering switches, reversing the sign of $\frac{dv}{dt}$, opening the door for disease resurgence through vaccine refusal. Of course, two other mechanisms can drive fear of vaccine to exceed fear of disease. One is an accumulation of adverse vaccine events represented by the $A$ compartment. Another mechanism (not included here) would be exogenous suppression of disease fear ($S_{fd}$) through statements by officials underestimating the threat.   
 
We turn now to the core scenarios of the model. Again, all numerical assumptions are provided in the text or the SI.

%%%%%%%%%%%%%%%%%%%%%%%%%%%%%%%%%%%%%%%%%%%%%%%%%%%%%%%%%%%%%%%%%%%
%%%%%%%%%%%%%%%%%%% SECTION 3 - THE RESULTS %%%%%%%%%%%%%%%%%%%%%%%%%%%%%%%%
%%%%%%%%%%%%%%%%%%%%%%%%%%%%%%%%%%%%%%%%%%%%%%%%%%%%%%%%%%%%%%%%
\section{Results}\label{results}
%%%%%%%%%%%%%%%%%%%   BASE SCENARIOS %%%%%%%%%%%%%%%%%%%
\subsection{Base scenarios}\label{baseScenarios}

%%%%%%%%%%%%%%%%%%%  SCENARIO 1  %%%%%%%%%%%%%%%%%%%
\subsubsection{Scenario 1: Contagious Disease Only}\label{baseScenario1}
\begin{figure}[h!]
\centering
\includegraphics[scale=0.33]{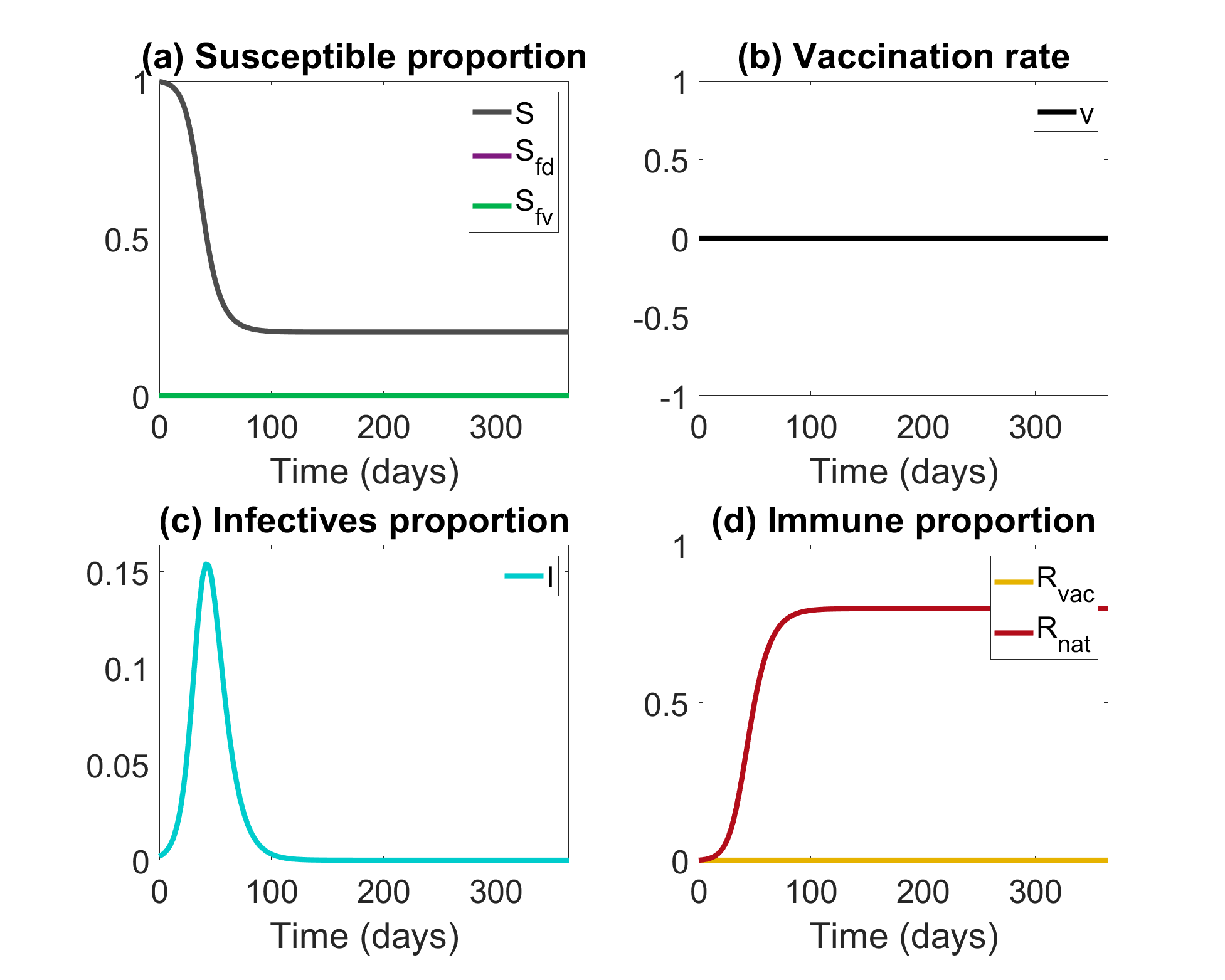}
\caption{Plots for Scenario 1 (contagious disease only): (a) The proportions of susceptibles without fear ($S$), disease fearful susceptibles ($S_{fd}$), and vaccine fearful susceptibles ($S_{fv}$). (b) Vaccination rate ($v$). (c) The proportion of infectives ($I$). (d) The proportion of recovered ($R_{nat}$) and vaccinated ($R_{vac}$) individuals. Note that about 80\% of the population become infected with the disease.}
\label{fig_scenario1}
\end{figure}

Here, we ``dock" the model to the classic case, an SIR epidemic with no fears, with a disease transmission rate $\beta$, and a single recovery (and subsequently immune) rate, $\gamma$.  In this case, Eqs (\ref{diffeq1}), (\ref{diffeq4}), and (\ref{diffeq5}) reduce to the Kermack-McKendrick model. A reference plot of the main dynamics is given in Figure \ref{fig_scenario1}, which illustrates our graphical strategy. To reduce clutter, it will prove useful to have four plots focused on different aspects of the coupled contagions: susceptibles, vaccine uptake, infection, and removals, as shown in Figure \ref{fig_scenario1}.

%%%%%%%  FIGURE 2 - no fears

With all fears and all vaccinations clamped at zero (see purple and green curves in Figure \ref{fig_scenario1}a  and black curve in Figure \ref{fig_scenario1}b), we see the classical one peak curve of infectives in Figure \ref{fig_scenario1}c, the falling susceptible curve in \ref{fig_scenario1}a, and the familiar recovered curve in Figure \ref{fig_scenario1}d. 

%%%%%%%%%%%%%%%%%%%  SCENARIO 2  %%%%%%%%%%%%%%%%%%%
\subsubsection{Scenario 2: Contagious Disease + Fear of Disease}\label{baseScenario2}
\begin{figure}[h!]
\centering
\includegraphics[scale=0.33]{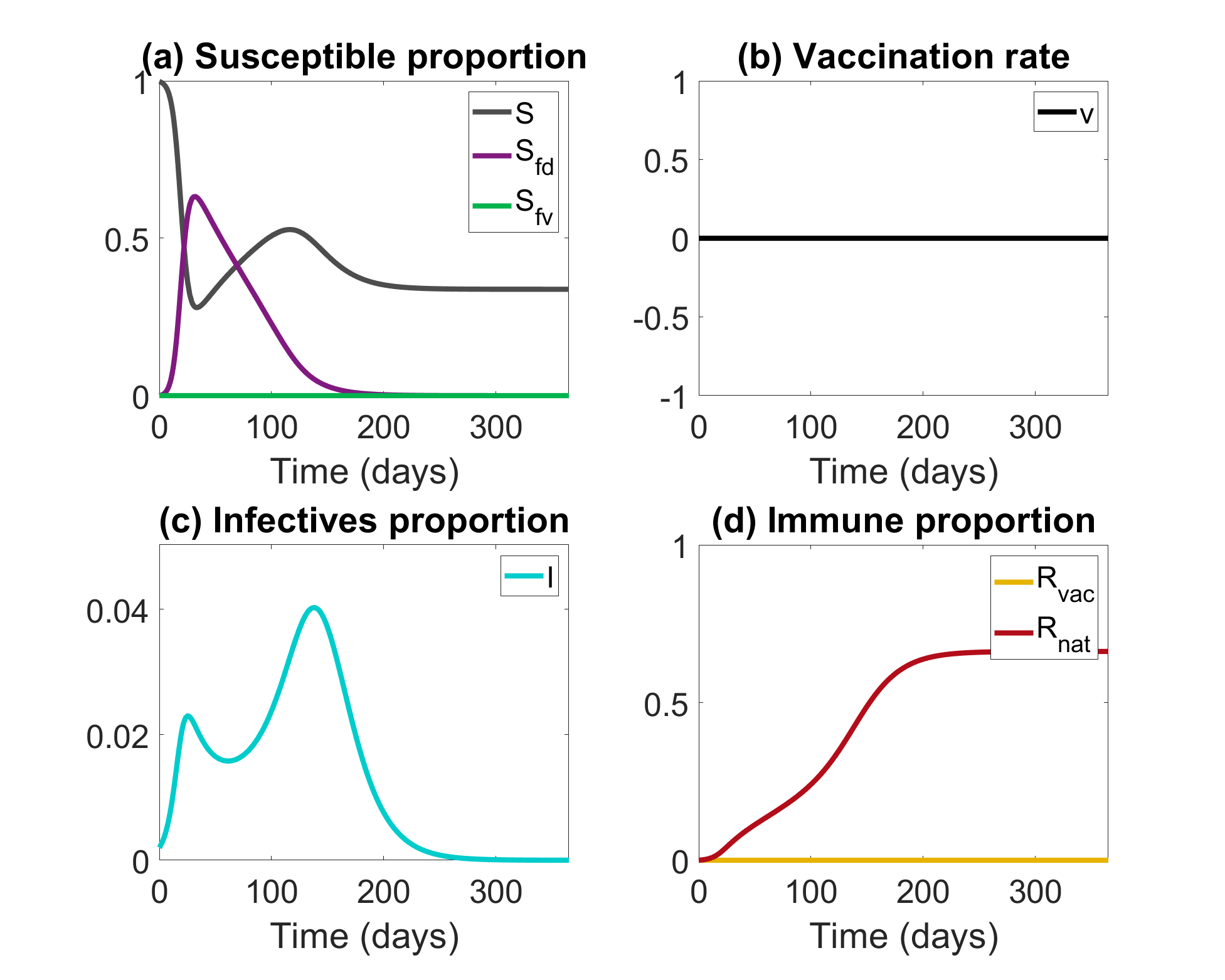}
\caption{Plots for Scenario 2 (Contagious disease + fear of the disease) (a)  The proportions of susceptibles without fear ($S$), disease fearful susceptibles ($S_{fd}$), and vaccine fearful susceptibles ($S_{fv}$). (b) Vaccination rate ($v$). (c) The proportion of infectives ($I$). (d) The proportion of recovered ($R_{nat}$) and vaccinated ($R_{vac}$) individuals. Note that about 66\% of the population become infected with the disease.}
\label{fig_scenario2}
\end{figure}

Now we add contagious fear of the pathogen, so there are two contagions, as depicted in Figure \ref{fig_scenario2}. The core narrative here is that the initial spike of infections (the blue curve) stimulates a fear spike (the purple curve). People reduce their contacts out of fear (this is modeled through $p$), which suppresses disease spread. As the disease wanes, however, so does the fear of it. Now, susceptibles go back into circulation, which pours fuel on the infective embers, and a second wave ensues. The second wave is larger than the first. Why? Because in our model, there are two mechanisms of fear decay, and they amplify one another. One mechanism is the "natural decay" governed by the parameter $\gamma_f$. The second is the "contagious," fear-reversal mechanism. People who have recovered from the disease are in contact with those who are still fearful. The recovereds' low fear is also transmitted, emboldening the fearful people in hiding to come 'out of the basement' when it is still unsafe. This 'complacency contagion,' if you will, amplifies the natural fear decay rate to produce a very sharp fear reduction. This pours a larger number of susceptibles onto the circulating infectives than would either fear decay mechanism alone. The result is that the second wave of the disease can be larger than the first, as occurred in 1918 \cite{he2011a}.  All of this is chronicled in Figure \ref{fig_scenario2}. Its robustness is explored in Section \ref{sensAnalysis}.

%%%%%%%  FIGURE 3 -- one fear 

\newpage
As the data science of social media shows \cite{depoux2020}, fear can spread much faster and much farther than the disease itself (a good thing when it induces preventive measures). For the earlier 2008 model, an analytic expression for the $R_0$ of fear, and conditions for fear of disease to spread faster than the disease itself, are given in \cite{epstein2008a}. The mathematics are different here and several analytical growth conditions for the present model are given in the SI text. An obvious reason for fear to outpace disease is that disease transmission requires direct physical contact while fear transmission does not. Indeed, there are two channels to acquire disease fear in our model--through contact with an infectious person (in the I compartment) or contact with a frightened susceptible person, in the $S_{fd}$ compartment. Scared individuals–whether sick or not–remove themselves from circulation, social distancing with an effectiveness governed by the parameter $p$. This endogenously affects the contact dynamic, and thus the disease epidemic itself. Sometimes, the self-isolation is sufficient to produce herd immunity and epidemic fade-out. (See Section \ref{sensAnalysis}).  In other cases, because disease prevalence is low, individuals recover from fear at a rate $\alpha_f$ despite the presence of disease. This releases fresh susceptibles onto the still-circulating infectives, generating a second wave, as shown in Figure \ref{fig_scenario2}. We now extend the model further, adding vaccination, but not yet the fear of it.

%%%%%%%%%%%%%%%%%%%  SCENARIO 3  %%%%%%%%%%%%%%%%%%%
\subsubsection{Scenario 3: Contagious Disease + Fear of Disease + Vaccinations}\label{baseScenario3}
 Vaccination can mitigate the second wave generated in Scenario 2, a beneficial result from a public health perspective.

%%%%%%%  FIGURE 4 -- one fear + vaccine
\begin{figure}[h!]
\centering
\includegraphics[scale=0.33]{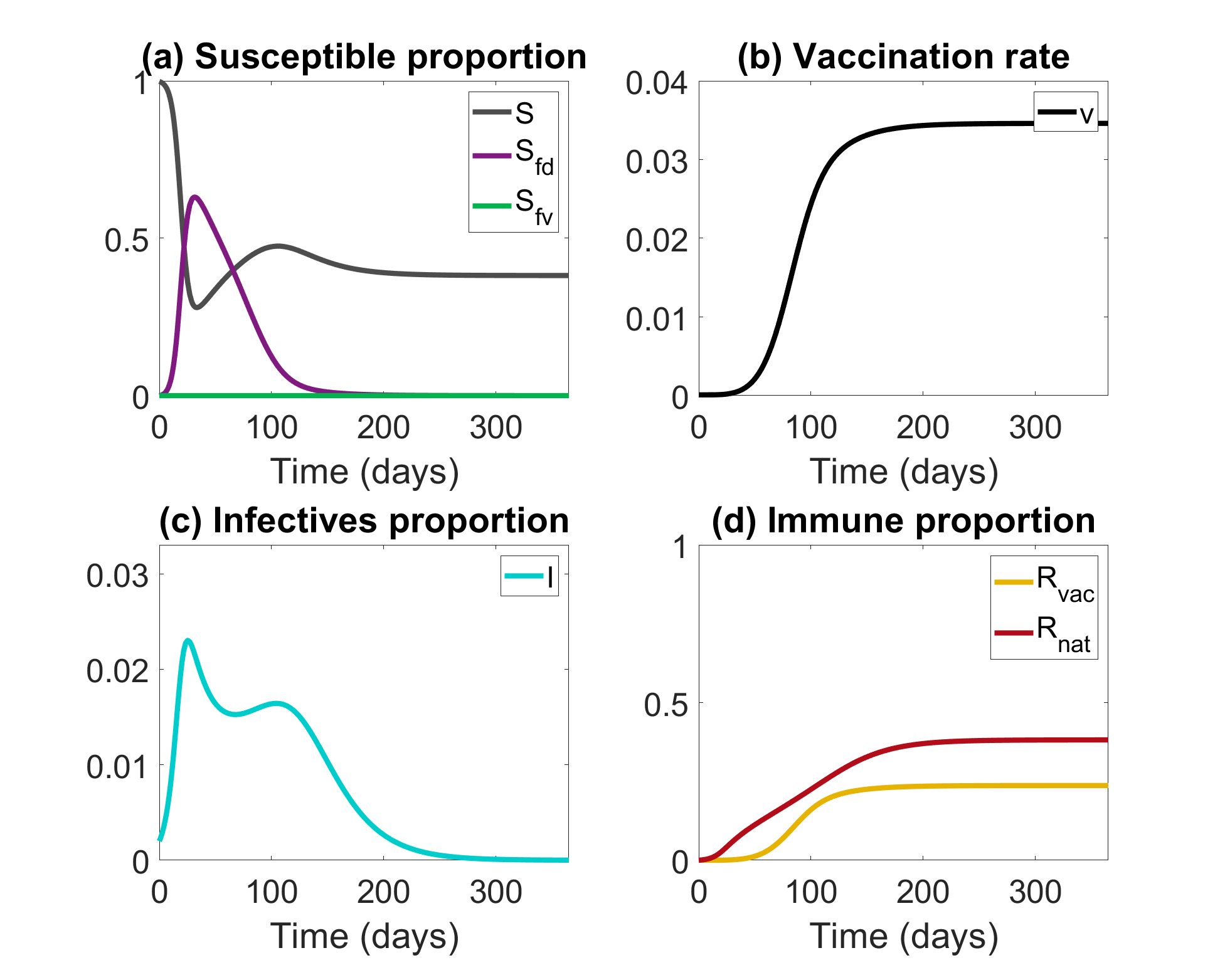}
\caption{Plots for Scenario 3 (Contagious disease + fear of the disease + vaccinations) (a)  The proportions of susceptibles without fear ($S$), disease fearful susceptibles ($S_{fd}$), and vaccine fearful susceptibles ($S_{fv}$). (b) Vaccination rate ($v$). (c) The proportion of infectives ($I$). (d) The proportion of recovered ($R_{nat}$) and vaccinated ($R_{vac}$) individuals. Note that about 38\% of the population become infected with the disease.}
\label{fig_scenario3}
\end{figure}

In Figure \ref{fig_scenario3}a, fear of disease (purple) exceeds fear of vaccine (green), which is clamped at zero. Vaccine uptake thus increases, as shown in Figure \ref{fig_scenario3}b. Now we see both "natural" and vaccine-induced removals (Figure \ref{fig_scenario3}d). The combined effect is to suppress the second wave, as evident from the lower-left infection curve. 

\newpage
Notice that, while the second wave of Figure \ref{fig_scenario3} is clearly suppressed, there is still a small second wave. The mechanism for the two peaks here lies in the assumed effectiveness of social distancing. We have assumed that the probability $p$ of contracting the disease while fearful is low. If we increase $p$, the peaks get closer together until (in the limit) they converge to a single peak again. The sensitivity of the phenomenon to variations in $p$ is given in Section \ref{sensAnalysis} below.

%%%%%%%%%%%%%%%%%%%  SCENARIO 4  %%%%%%%%%%%%%%%%%%%
\subsubsection{Scenario 4: Contagious Disease + Fear of Disease + Vaccinations + Fear of Vaccinations}\label{baseScenario4}
In Scenario 4, the fear of vaccination "wins," and the outbreak is again unmitigated. People do vaccinate at the beginning of the outbreak but stop too soon because the fear ordering reverses. 

%%%%%%%  FIGURE 5 -- two fears + vaccine
\begin{figure}[h!]
\centering
\includegraphics[scale=0.33]{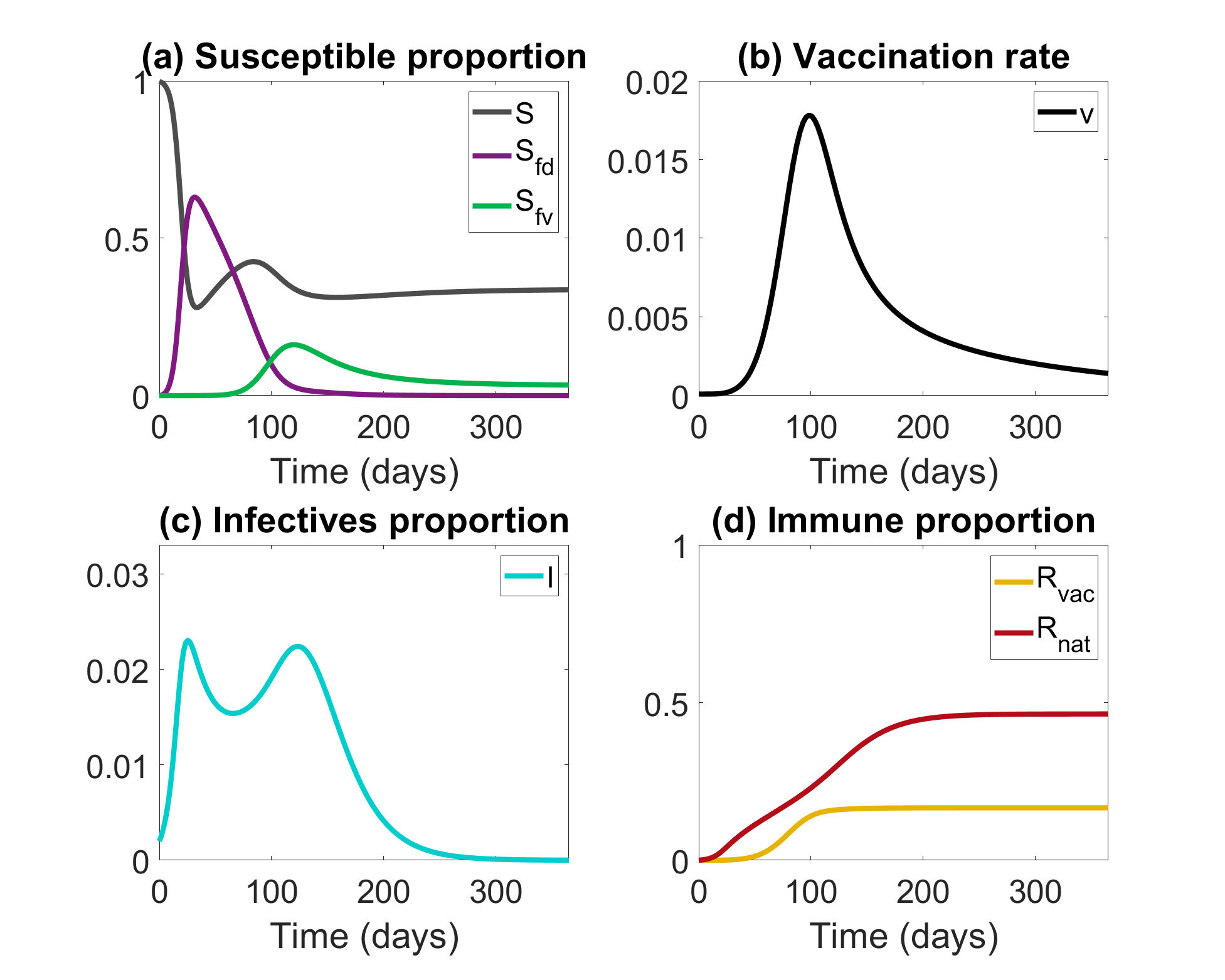}
\caption{Plots for Scenario 4 (Contagious disease + fear of the disease + vaccinations + fear of the vaccinations (a)  The proportions of susceptibles without fear ($S$), disease fearful susceptibles ($S_{fd}$), and vaccine fearful susceptibles ($S_{fv}$). (b) Vaccination rate ($v$). (c) The proportion of infectives ($I$). (d) The proportion of recovered ($R_{nat}$) and vaccinated ($R_{vac}$) individuals. Note that about 46\% of the population become infected with the disease.}
\label{fig_scenario4}
\end{figure}

In Figure \ref{fig_scenario4}a, we see that the fear ordering changes at roughly 100 days, at which point fear of vaccine (green) rises above the fear of disease (purple). This reverses the sign of the rate of change of $v(t)$ [(Eq \ref{diffeq8})], and a second wave of infections ensues.  

The base scenarios exhibit several mechanisms for the emergence, timing, size, and decay of multiple waves. We now explore their sensitivity to various parameters. 
%\clearpage % Forcing all the figures we defined to be printed before this point

%%%%%%%%%%%%%%%%%%%%%%%%%%%%%%%%%%%%%%%%%%%%%%%%%%%%%%%%%%%%%%%%%%%
%%%%%%%%%%%%%%%% SECTION 4 - SENSITIVITY ANALYSIS %%%%%%%%%%%%%%%%%%%%%%%%%%%%%%%%
%%%%%%%%%%%%%%%%%%%%%%%%%%%%%%%%%%%%%%%%%%%%%%%%%%%%%%%%%%%%%%%%
\newpage
\section{Sensitivity Analysis}\label{sensAnalysis}

\subsection {One Fear (Scenario 2)}\label{oneFear}

%%%%%%%%%%%%%%%%%%% RELATIVE RISK, P  4.1 %%%%%%%%%%%%%%%%%%%
\subsubsection{Sensitivity to $p$, the relative risk reduction due to protective behaviors}\label{changingP}
\begin{figure}[h!]
\centering
\includegraphics[scale=0.35]{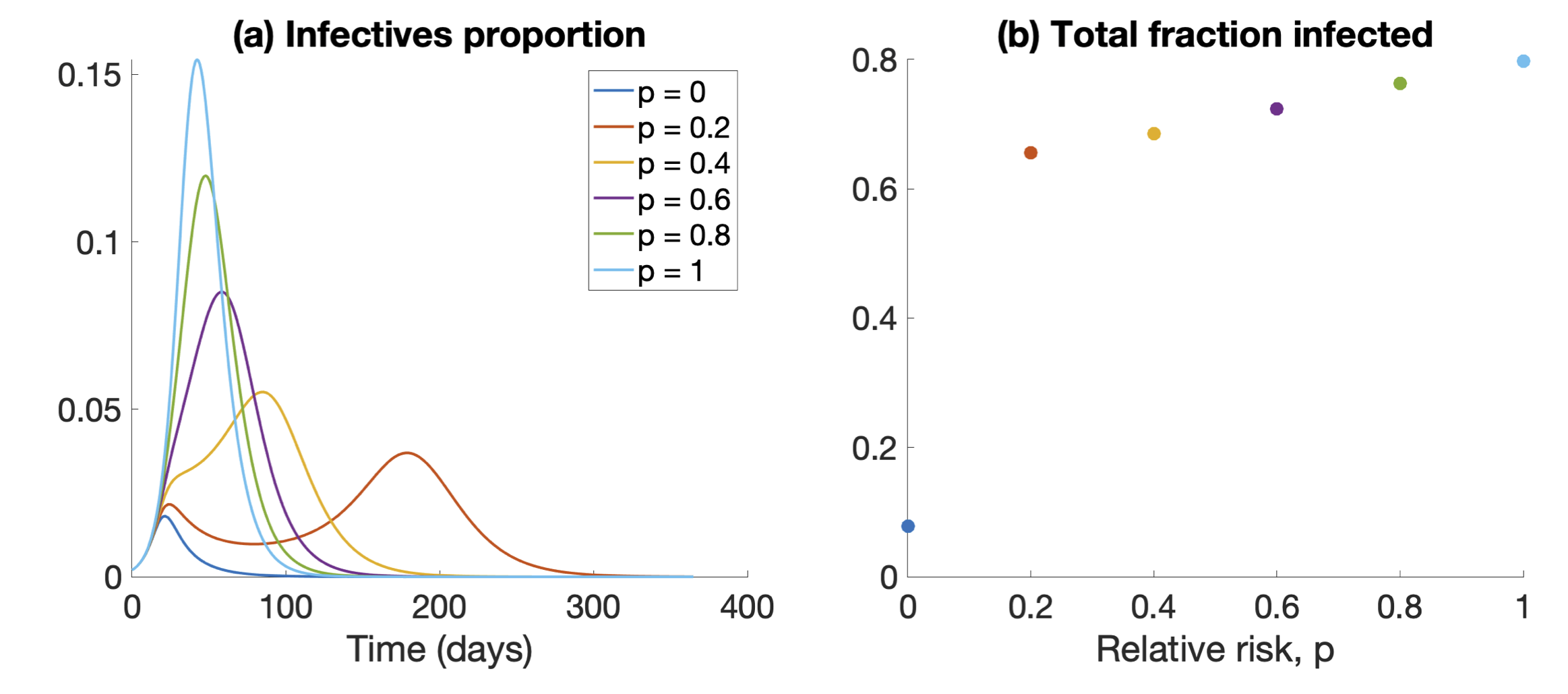}
\caption{The effect of changing the relative risk, $p$, on disease spread. (a) the proportion of infectives ($I$) vs. time. (b) the total fraction of the population that contacts the disease. All parameters other than $p$ are as in Scenario 2 (Table S1, column 2).}
\label{fig_oneFear_changeP}
\end{figure}

To begin, we return to the case of the disease and fear of the disease only  (Scenario 2, Section \ref{baseScenario2}) and study the effect of changing the relative risk $p$ of acquiring the disease for pathogen-fearful individuals. Figure \ref{fig_oneFear_changeP} shows that if $p$ is decreased to 0, meaning that those who are fearful of the disease go into hiding and have a 0\% chance of contracting the disease, the epidemic will be prevented (we see that only about 10\% of the population gets the disease). By contrast, as $p$ increases, fearful individuals become more risk-neutral (increasing their likelihood of contracting the pathogen), and the epidemic worsens; at its worst, with the fearful individuals not altering their behavior at all ($p=1$), we see that about 80\% of the population becomes infected. 

%%%%%%%  FIGURE 6 -- one fear, changing p

We see also that there are single-wave and two-wave regimes, depending on $p$. If $p$ is small (less than about 0.4) but positive, we see a second wave emerge as fearful individuals hide away and then return to circulation. If $p$ is larger than about 0.4, then the fearful individuals don’t lower their risk enough to preserve a susceptible population sufficient to produce a second wave.

\newpage
%%%%%%%%%%%%%%%%%%%  RATE OF CONTAGIOUS FEAR LOSS, ALPHA  4.2 %%%%%%%%%%%%%%%%%%%
\subsubsection{Sensitivity to $\alpha_f$, the effective contact rate of fear-loss}\label{changingAlphaF}
\begin{figure}[h!]
\centering
\includegraphics[scale=0.8]{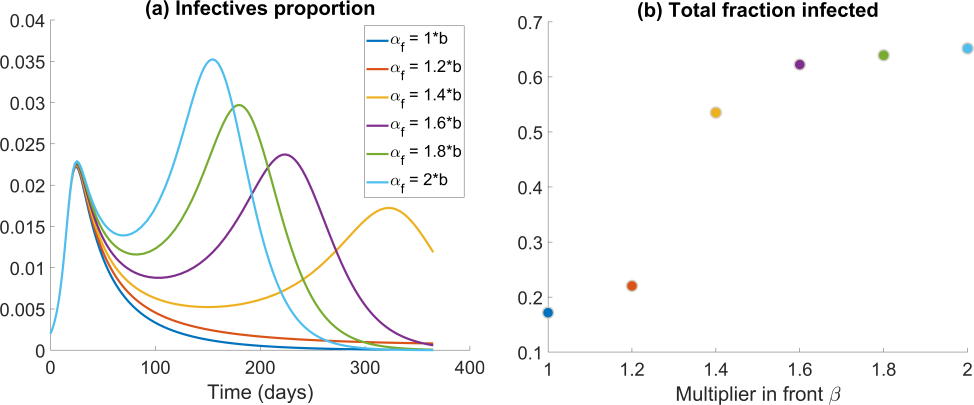}
\caption{The effect of the effective contact rate of fear loss $\alpha_f$ on the (a)  proportion of infectives ($I$) vs. time., and the (b)  total fraction of the population that contacts the disease. All parameters other than $\alpha_f$ remain as in Table S1, column 2.}
\label{fig_oneFear_changeAlpha}
\end{figure}

As noted earlier, an important extension (among several) of the original coupled contagion model \cite{epstein2008a} is our inclusion of a second mechanism of fear loss. In addition to spontaneous loss of fear, disease fearful persons ($S_{fd}$) may lose their fear by interacting with recovered persons ($R_{nat}$). The effective contact rate for this interaction is $\alpha_f$. We can see that the infectives curve of different $\alpha_f$ values overlap until the first peak is reached; see Figure \ref{fig_oneFear_changeAlpha}a. The curves do not differ because the number of recovered in the beginning is too low to reduce fear significantly. The differences become apparent once the infectives curve drops. Higher values of $\alpha_f$ cause people to lose their fear and abandon their protective measures. This process increases the number of persons that are infected and in turn, the number of recovered. The larger number of recovered causes a larger fraction of persons to lose their fear of the disease and so on. The result is a second wave when the contact rate is sufficiently high. As $\alpha_f$ increases, the second peak is higher and occurs sooner. This process increases the fraction of infected persons; see Figure \ref{fig_oneFear_changeAlpha}b.

%%%%%%%  FIGURE 7 -- one fear, changing alpha_f

\newpage
%%%%%%%%%%%%%%%%%%%  TWO FEARS 4.3 %%%%%%%%%%%%%%%%%%%
\subsection{Two fears (Scenario 4)}\label{twoFears}
\subsubsection{Sensitivity to $\beta_{fd}$ and $\beta_{fv}$, the fear contact rates}\label{noOfPeaks}
\begin{figure}[h!]
	\centering
	\includegraphics[scale=0.32]{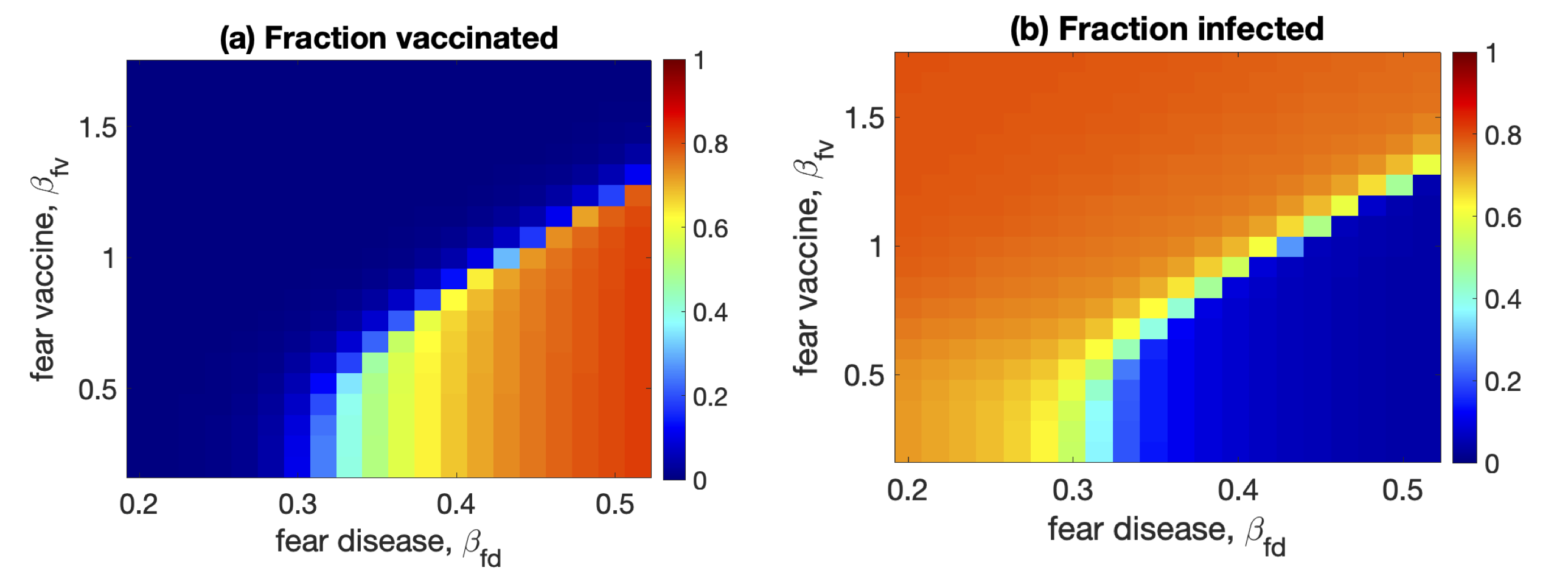}
	\caption{The effect of changing the fears contact rates  ($\beta_{fd}$,$\beta_{fv}$) on the (a) fraction of the population vaccinated and (b) infected with the disease.}
	\label{fig_twoFears_changeBeta_f_heat}
\end{figure}

We now focus on the two-fears scenario (Scenario 4) and explore how the contact rate of the fear of the disease ($\beta_{fd}$) and the contact rate of the fear of the vaccine ($\beta_{fv}$) influence the model behavior. The two effective contact rates determine how fast the fears are transmitted in the population. The fraction of vaccinated and infected persons as a function of ($\beta_{fd}$, $\beta_{fv}$) is shown in Figure \ref{fig_twoFears_changeBeta_f_heat}. When fear of the vaccine is transmitted sufficiently faster than the fear of the disease (dark blue region in Figure \ref{fig_twoFears_changeBeta_f_heat}a), the population eschews vaccine, and a large portion of the population becomes infected with the disease (over 75\%, as shown in bright orange region of Figure \ref{fig_twoFears_changeBeta_f_heat}b). When the fear the disease is transmitted fast with a sufficiently low transmission rate of the fear of the vaccine, a small proportion of the population is infected (the blue region in \ref{fig_twoFears_changeBeta_f_heat}b). The fear of the disease spreads too slowly, and the number of persons who adopt protective behavior is too low to create two peaks.

\begin{figure}[h!]
\centering
\includegraphics[scale=0.32]{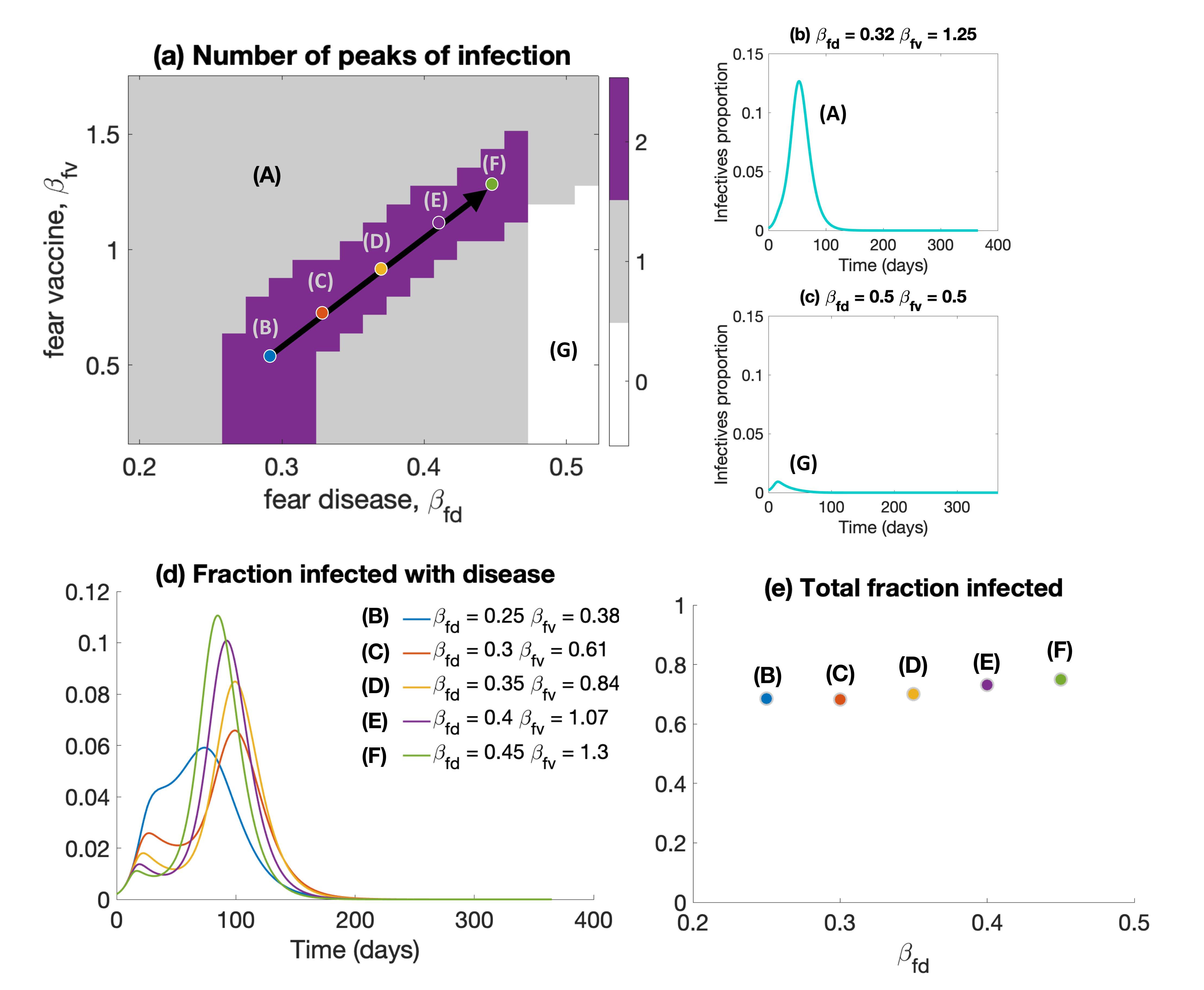}
\caption{The effect of changing the fears' contact rates  ($\beta_{fd}$, $\beta_{fv}$): (a) The number of peaks in the infectives curve as a function of the two contact rates. (b) the infectives curve for a case of one peak (c) the infectives curve for a case of no peak (no outbreak). (d) five infectives curves with two peaks. (e) the total number of infected persons for each of the five cases.}
\label{fig_twoFears_changeBeta_f}
\end{figure}

We now explore how the fears' contact rates ($\beta_{fd}$, $\beta_{fv}$) affect the number of peaks that occur in the infectives curve (I). We define a peak as a local maximum with a proportion of infectives above 0.01.  Figure \ref{fig_twoFears_changeBeta_f}a shows that the model produces zero, one, or two peaks. Point A is located in a region of fast transmission of the vaccine fear, which leads to few vaccinated individuals. Only one peak is generated (see Figure \ref{fig_twoFears_changeBeta_f}b) because the disease's level of fear is too low for temporarily decreasing the disease spread, and a majority of the population is infected.

Two peaks are generated in the purple area of Figure \ref{fig_twoFears_changeBeta_f}a. The two peaks region in ($\beta_{fd},\beta_{fv}$) space can be well-represented by the equation $\beta_{fv} = 4.6(\beta_{fd}-0.45)+1.3$ (see black arrow). Notice that this area includes the interface between the red and blue regions of Figure \ref{fig_twoFears_changeBeta_f_heat} which indicate the transition between the high and low fractions of vaccinated and infected persons. As we move along this arrow toward larger values of $\beta_{fd}$ and $\beta_{fv}$, we see that the second peak in the infection increases while the first peak decreases (Figure \ref{fig_twoFears_changeBeta_f}d). As more fear enters the population, we see individuals fearful of the disease hiding out (decreasing the first peak of infection), driving the vaccination rate up, and then a disproportionate fear of the vaccine (increasing the second peak). Along this arrow, the total fraction of the population that becomes infected remains relatively constant (Figure \ref{fig_twoFears_changeBeta_f}e) as the arrow is parallel to the interface of the two regions.

If $\beta_{fd}$ is high enough (above $\sim$0.45 per day), there will be no epidemic unless $\beta_{fv}$ is sufficiently high. Even then, there will be only one peak because there are too few susceptibles to cause an initial peak high enough to reverse the order of fears (that is, too many are fearful of the disease, reducing their transmission rate); see Figure \ref{fig_twoFears_changeBeta_f}c.

%%%%%%%  FIGURE 8 -- two fears, changing infectivity rate -- heat maps

%%%%%%%  FIGURE 9 -- two fears, changing infectivity rate

\newpage
\subsubsection{Sensitivity to $\sigma$, the fraction of adverse reactions}\label{adverseReactions}
\begin{figure}[h!]
\centering
\includegraphics[width = 0.95\textwidth]{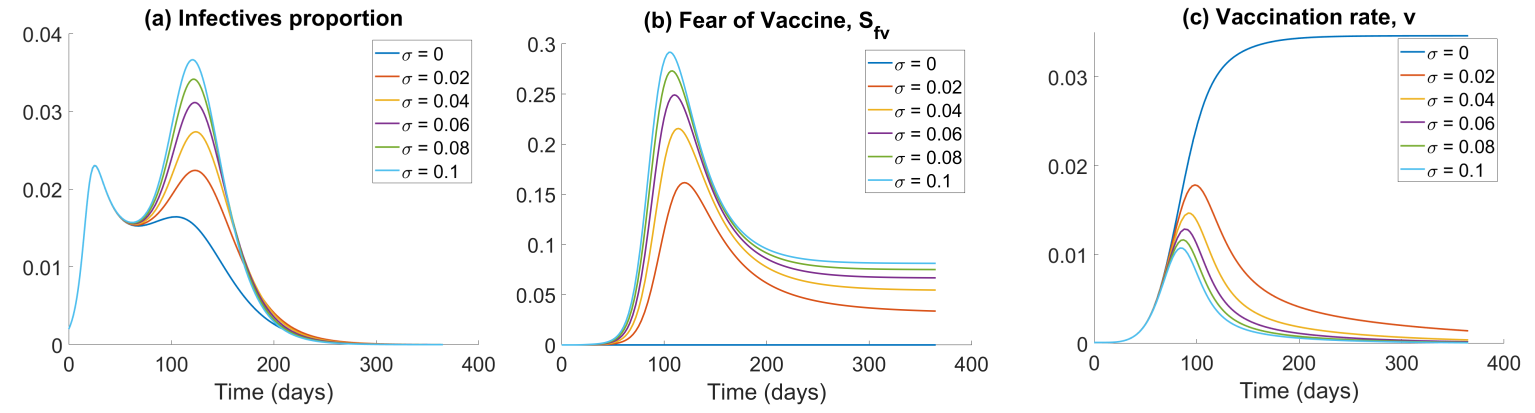}
\caption{The effect of changing the fraction of adverse effects from vaccinations ($\sigma$) on the (a) proportion of infectives, (b) the proportion of susceptibles that fear the vaccine, and (c) the vaccination rate.}
\label{fig_twoFears_changeSigma}
\end{figure}

A proportion $\sigma$ of individuals experience adverse reactions to the vaccine and develop a temporary fear of the vaccine, which may be transmitted to susceptible individuals. The more people that have such reactions, the easier it is for the fear to spread.  When we increase the proportion of adverse reactions, the second peak in the proportion of infectives rises while the first peak remains unaffected (Figure \ref{fig_twoFears_changeSigma}a). The first peak remains the same because too few persons fear the vaccine (Figure \ref{fig_twoFears_changeSigma}b), and they have a negligible effect on the vaccination rate (Figure \ref{fig_twoFears_changeSigma}c). After the first peak,  the fear of the vaccine spreads, the vaccination rate drops, and more people become infected, leading to the second peak.

%%%%%%%  FIGURE 11 -- two fears, changing infectivity rate - increasing along two peaks

%%%%%%%  FIGURE 12 -- two fears, changing infectivity rate - increasing along two peaks

%\begin{figure}[h!]
%\centering
%\includegraphics[width = 0.95\textwidth]{twoFears_changingEta.png}
%\caption{ Effect of changing the fraction of those vaccinated that have adverse %effects, $\eta$. }
%\label{fig_twoFears_changeBeta_eta}
%\end{figure}

\newpage
Further sensitivity analyses can, of course, be conducted. But these demonstrate how the scenario dynamics respond to variations in several key parameters. The SI gives parameter values and initial conditions for the scenarios and derives analytical expressions for the growth (the $R_n$ values) of the disease and fear epidemics.  The last of these specialize to give the condition for a fear epidemic in the absence of disease, which is also shown.

%%%%%%%%%%%%%%%%%%%%%%%%%%%%%%%%%%%%%%%%%%%%%%%%%%%%%%%%%%%%%%%%%%%
%%%%%%%%%%%%%%%% SECTION 5 - CONCLUSIONS  %%%%%%%%%%%%%%%%%%%%%%%%%%%%%%%%
%%%%%%%%%%%%%%%%%%%%%%%%%%%%%%%%%%%%%%%%%%%%%%%%%%%%%%%%%%%%%%%%

\section{Conclusions}\label{conclusion}
We have extended earlier work on coupled contagion dynamics \cite{epstein2008a}. In addition to a contagious disease and contagious fear of it, we have added a second contagion: fear of the control, in this case, vaccine. In addition, unlike \cite{epstein2008a}, we include both the classical extinction of fear and its contagious evaporation. The interaction of these entangled contagions reveals several novel behavioral mechanisms for multiple waves of infection and for their timing, size, and form. To adequately capture such mechanisms, infectious disease modeling must begin to incorporate behavioral neuroscience. Human behavior is complex and involves interacting affective, deliberative, and social components. To be sure, some health decisions qualify as canonically rational. But often, as Hume noted, ``Reason is a slave to the passions,” the more so in settings of extreme stress like pandemics. Simple models grounded in the neuroscience of fear and its transmission can deepen epidemic modeling.

\newpage

\newpage
%\bibliography{FearManuscript_bib.bib}

\end{document}

%%%%%%%%%%%%%%%%%%%%%%%%%%%%%%%%%%%%%%%%%%%%%%%%%%%%%%%%%%%%%%%%%%%
%%%%%%%%%%%%%%%% SUPPLEMENTARY INFO  %%%%%%%%%%%%%%%%%%%%%%%%%%%%%%%%
%%%%%%%%%%%%%%%%%%%%%%%%%%%%%%%%%%%%%%%%%%%%%%%%%%%%%%%%%%%%%%%%
\section*{Supplementary Information}\label{SI}

%%%%%%%%%%%%%%%%%%%  S1 PARAMETERS %%%%%%%%%%%%%%%%%%%
\subsection*{Parameters and initial conditions}\label{Params}
The parameters employed in each scenario are shown in Table \ref{S1_Table} below, illustrating again the cumulative nature of the exercise.

\begin{table}[h!]
\renewcommand{\arraystretch}{1.2}
\begin{tabular}{|c|c|c|c|c|}
\hline
\cellcolor[gray]{0.8}  \textbf{Parameter} &  \textbf{Scenario 1:} &  \cellcolor[gray]{0.9}  \textbf{Scenario 2:} &   \textbf{Scenario 3:} &  \cellcolor[gray]{0.9}  \textbf{Scenario 4: }  \\
\cellcolor[gray]{0.8}  \textbf{ } &  \textbf{no fear} &  \cellcolor[gray]{0.9}  \textbf{one fear} &    \textbf{one fear + vaccine } &  \cellcolor[gray]{0.9}  \textbf{two fears + vaccine }  \\
\hline 
\cellcolor[gray]{0.8}$\beta$ & 2/7 & \cellcolor[gray]{0.9}2/7 & 2/7 &\cellcolor[gray]{0.9} 2/7 \\
\hline
\cellcolor[gray]{0.8}$\gamma$ & $\beta/2$ & \cellcolor[gray]{0.9}$\beta/2$ & $\beta/2$ & \cellcolor[gray]{0.9}$\beta/2$\\
\hline
\cellcolor[gray]{0.8}$\beta_{fd}$ &  0 & \cellcolor[gray]{0.9}$1.1\beta$ & $1.1\beta$ & \cellcolor[gray]{0.9}$1.1\beta$\\
\hline
\cellcolor[gray]{0.8}$\alpha_{f}$ & 0 & \cellcolor[gray]{0.9}$2.2\beta$  & $2.2\beta$ & \cellcolor[gray]{0.9}$2.2\beta$\\
\hline
\cellcolor[gray]{0.8}$\gamma_f$ & 0 & \cellcolor[gray]{0.9}0.05 & 0.05 & \cellcolor[gray]{0.9}0.05\\
\hline
\cellcolor[gray]{0.8}$p$ & 0 & \cellcolor[gray]{0.9}0.25 & 0.25 & \cellcolor[gray]{0.9}0.25\\
\hline
\cellcolor[gray]{0.8}$\eta$ & 0 & \cellcolor[gray]{0.9}0& 0.8 & \cellcolor[gray]{0.9}0.8\\
\hline
\cellcolor[gray]{0.8}$\sigma$ & 0 & \cellcolor[gray]{0.9}0 & 0.02 &\cellcolor[gray]{0.9}0.02 \\
\hline
\cellcolor[gray]{0.8}$\epsilon$ & 0 & \cellcolor[gray]{0.9}0 & 0.2 & \cellcolor[gray]{0.9}0.2\\
\hline
\cellcolor[gray]{0.8}$\beta_{fv}$ & 0 & \cellcolor[gray]{0.9}0 & 0 & \cellcolor[gray]{0.9}$1.6\beta$ \\
\hline
\end{tabular}
\caption{Parameter values used in the scenarios  in Section 3.1.}
\label{S1_Table}
\end{table}

\begin{table}[h!]
\renewcommand{\arraystretch}{1.2}
\begin{tabular}{|c|c|c|c|c|}
\hline
\cellcolor[gray]{0.8}  \textbf{Variable} &  \textbf{Scenario 1:} &  \cellcolor[gray]{0.9}  \textbf{Scenario 2:} &   \textbf{Scenario 3:} &  \cellcolor[gray]{0.9}  \textbf{Scenario 4: }  \\
\cellcolor[gray]{0.8}  \textbf{ } &  \textbf{no fear} &  \cellcolor[gray]{0.9}  \textbf{one fear} &    \textbf{one fear + vaccine } &  \cellcolor[gray]{0.9}  \textbf{two fears + vaccine }  \\
\hline 
\cellcolor[gray]{0.8} $S(0)$ & 0 & \cellcolor[gray]{0.9}0 & 0 &\cellcolor[gray]{0.9} 0 \\
\hline
\cellcolor[gray]{0.8} $S_{fd}(0)$ & 0 & \cellcolor[gray]{0.9}0 & 0 & \cellcolor[gray]{0.9} 0\\
\hline
\cellcolor[gray]{0.8} $S_{fv}(0)$ &  0 & \cellcolor[gray]{0.9} 0 & 0 & \cellcolor[gray]{0.9} 0\\
\hline
\cellcolor[gray]{0.8}$I(0)$ & 0.002 & \cellcolor[gray]{0.9} 0.002  & 0.002 & \cellcolor[gray]{0.9} 0.002\\
\hline
\cellcolor[gray]{0.8}$R_{nat}(0)$ & 0 & \cellcolor[gray]{0.9}0 & 0 & \cellcolor[gray]{0.9}0\\
\hline
\cellcolor[gray]{0.8}$R_{vac}(0)$ & 0 & \cellcolor[gray]{0.9}0 & 0 & \cellcolor[gray]{0.9}0\\
\hline
\cellcolor[gray]{0.8}$A(0)$ & 0 & \cellcolor[gray]{0.9}0& 0 & \cellcolor[gray]{0.9} 0\\
\hline
\cellcolor[gray]{0.8}$v(0)$ & 0 & \cellcolor[gray]{0.9}0 & 0.0001 &\cellcolor[gray]{0.9} 0.0001 \\
\hline
\end{tabular}
\caption{Initial values used in the scenarios  in Section 3.1.}
\label{S2_Table}
\end{table}

%%%%%%%%%%%%%%%%%%%  S2 RN GROWTH CONDITIONS %%%%%%%%%%%%%%%%%%%
\subsection*{Growth conditions: $R_n$ values}\label{growthCond}

%%%%%%%%%%%%%%%%%%%  S2.1 RN NO FEAR %%%%%%%%%%%%%%%%%%%
\subsubsection*{Growth conditions for the no-fear model}

\begin{figure}[h!]
\centering
\includegraphics[width = 0.45\textwidth]{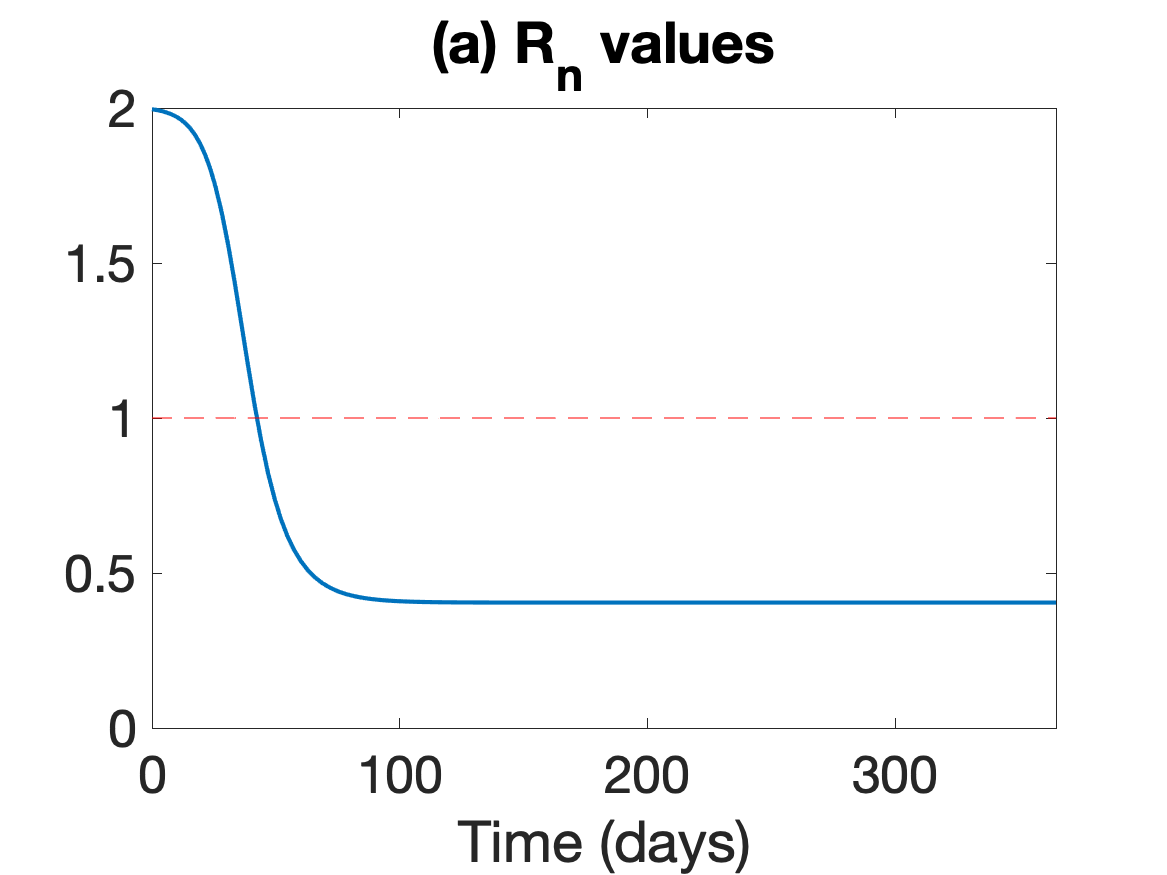}
\includegraphics[width = 0.45\textwidth]{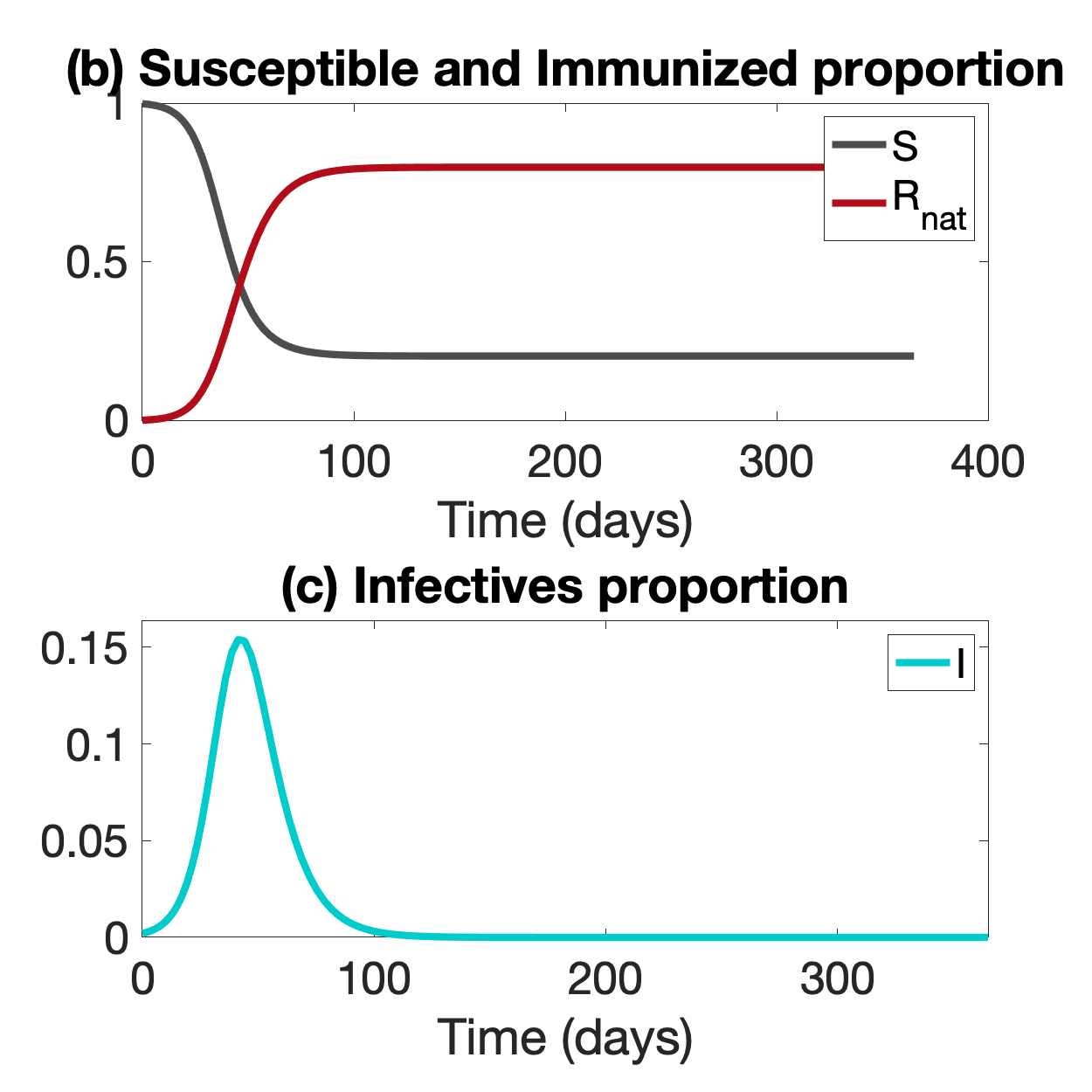}
\caption{Classic SIR model (Scenario 1). (a) The reproduction number ($R_n$) value for the no-fear case plotted over time. (b) The proportion of susceptible ($S$) and recovered ($R_{nat}$) individuals (c) The proportion of infectives ($I$).}
\label{fig_noFear_Rn}
\end{figure}

As shown in the text, in the absence of fear, our model reduces to the classical SIR model. The resulting $R_n$ values are no exception. Here, we calculate the condition for continued growth of an epidemic. 
\begin{align*}
\frac{dI}{dt} &> 0\\
\beta IS +p\beta I S_{fd} + \beta IS_{fv} - \gamma I &> 0\\
\beta\left( \frac{S+pS_{fd} + S_{fv} }{\gamma}\right) &> 1
\end{align*}
If we consider the case of no fear by setting $S_{fd} = S_{fv} = 0$, we arrive at the typical growth condition: $R_n = \frac{\beta S}{\gamma}>1$. If we plot this value for the simulation shown in Section 3.1.1, Figure 2, we see that when the $R_n$ value is above 1, the disease is increasing; see Figure \ref{fig_noFear_Rn}

%%%%%%%  FIGURE 13 --  no fears - Rn values

\subsubsection*{Growth conditions for a fear epidemic}
In a similar way, we can calculate an $R_n$ value for the spread of the fear of the disease by finding a condition on which the derivative of $S_{fd}$ is positive

\begin{align*}
\frac{dS_{fd}}{dt} &> 0\\
-p\beta I S_{fd} - v S_{fd} - \gamma_f S_{fd} - \alpha_f R_{nat}S_{fd} + \beta_{fd}S_{fd}S + \beta_{fd}IS &> 0\\
\beta_{fd}S_{fd}S + \beta_{fd}IS &> S_{fd}(p\beta I + v + \gamma_f + \alpha_f R_{nat})\\
\beta_{fd} \left( \frac{S + \frac{IS}{S_{fd}}}{p\beta I + v + \gamma_f + \alpha_f R_{nat}}  \right) &>1
\end{align*}

Finally, the extreme case is fear contagion in the absence of actual disease. Baseless fear contagions are common outside public health. The Salem Witch Trials of 1692 come immediately to mind.  But in public health there are purely psychogenic contagions like Morgellan’s disease, noted earlier. If we assume that there is no disease ($I = 0$), the growth condition above simplifies to 

\[ \frac{ \beta_{fd}S}{v + \gamma_f + \alpha_f R_{nat}}   >1\]

\begin{figure}[h!]
\centering
\includegraphics[width = 0.45\textwidth]{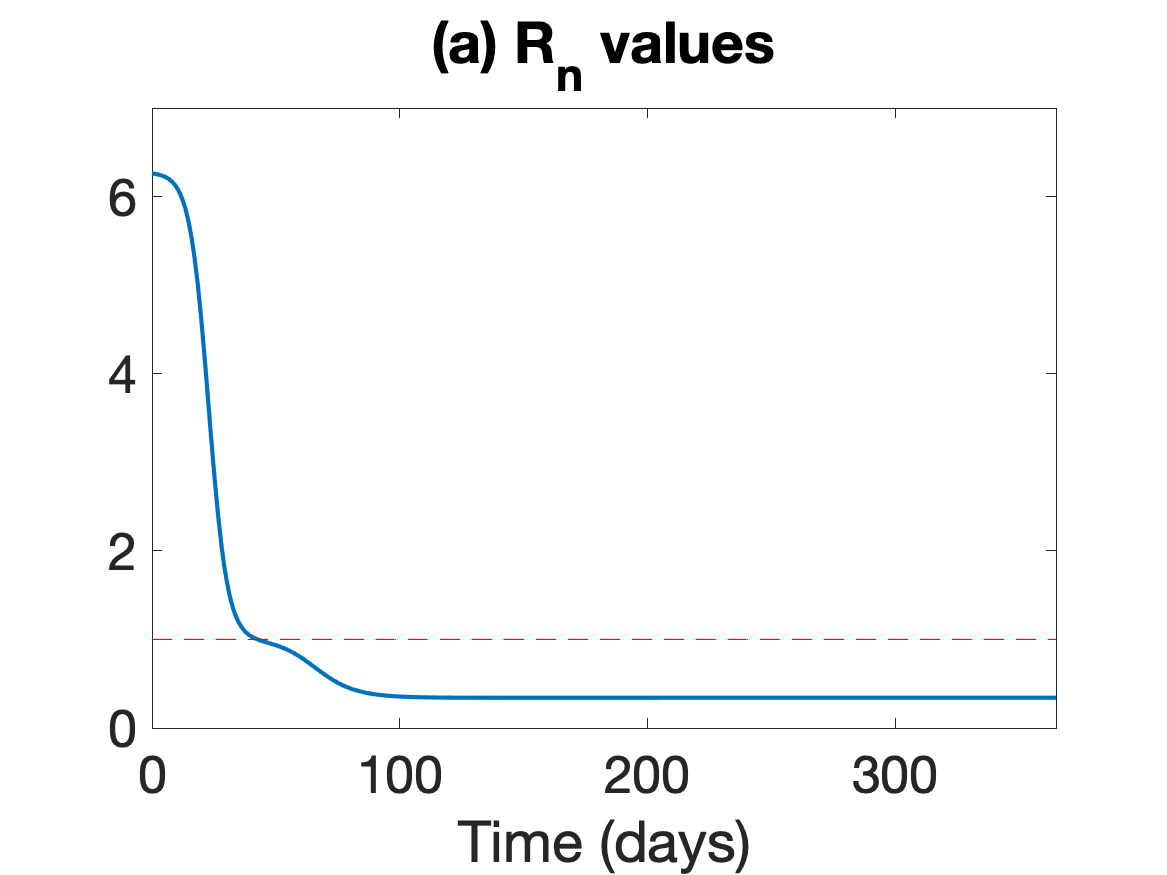}
\includegraphics[width = 0.45\textwidth]{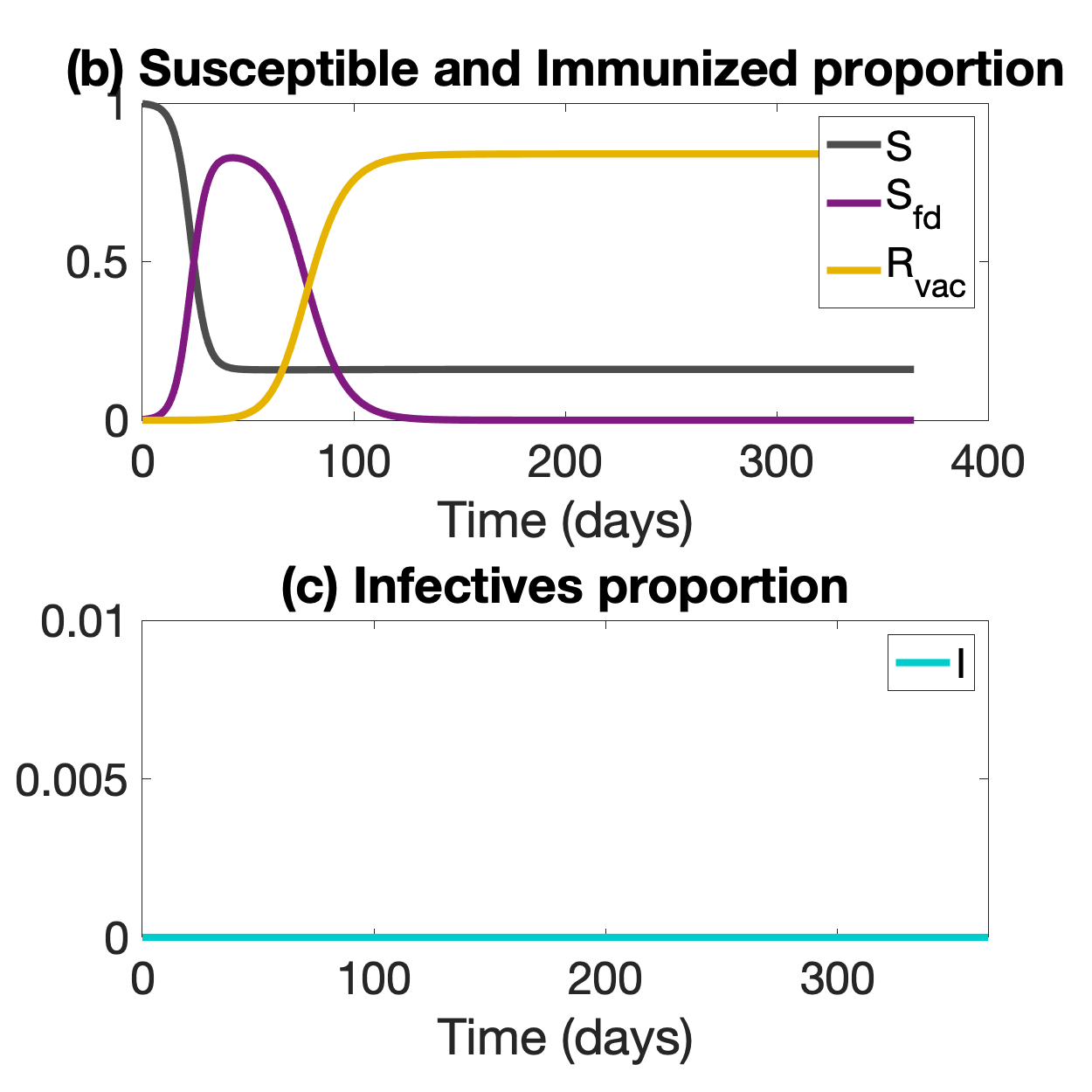}
\caption{Fear epidemic. (a) The reproduction number ($R_n$) value for the fear of the disease plotted over time. (b) The proportion of susceptible ($S$), disease fearful ($S_{fd}$) and vaccinated ($R_{vac}$) individuals. (c) The proportion of infectives ($I$).}
\label{fig_justFear_Rn}
\end{figure}

 \noindent Finally, if we begin with one baselessly fearful individual, we see that for values of disease-free $R_n$ greater than 1, we have an epidemic of fear without disease; see Figure \ref{fig_justFear_Rn}.

%%%%%%%  FIGURE 14 --  no disease - Rn values
